\documentclass[10pt,letterpaper]{article}
\usepackage[utf8]{inputenc}
\usepackage{amsmath}
\usepackage{amsfonts}
\usepackage{mathtools}
\usepackage{amssymb}
\usepackage{graphicx}
\usepackage{verbatim}
\graphicspath{{figures/}}
\usepackage{multicol}
\usepackage{authblk}
\usepackage[left=1.5cm,right=1.5cm,top=1.5cm,bottom=1.5cm]{geometry}
\usepackage{color}
\usepackage{fixltx2e}

\newcommand\ddfrac[2]{\frac{\displaystyle #1}{\displaystyle #2}}
\newcommand{\lap}{\mbox{$\cal L$}}
\newcommand{\ra}{\rightarrow}
\newcommand{\rla}{\rightleftharpoons}

\newcommand{\R}{\mathbb{R}}
\newcommand{\Z}{\mathcal Z}
\newcommand{\pr}{\mbox{Pr}}

\author[1]{John W. Biddle}
\author[1]{Jeremy Gunawardena}
\affil[1]{Department of Systems Biology, Harvard Medical School, Boston, United States}
\title{Reversal symmetries for cyclic paths away from thermodynamic equilibrium}
\begin{document}
\maketitle
\begin{abstract}
If a system is at thermodynamic equilibrium, an observer cannot tell whether a film of it is being played forward or in reverse: any transition will occur with the same frequency in the forward as in the reverse direction. However, if expenditure of energy changes the rate of even a single transition to yield a non-equilibrium steady state, such time-reversal symmetry undergoes a widespread breakdown, far beyond the point at which the energy is expended. An explosion of interdependency also arises, with steady-state probabilities of system states depending in a complicated manner on the rate of every transition in the system. Nevertheless, in the midst of this global non-equilibrium complexity, we find that cyclic paths have reversibility properties that remain local, and which can exhibit symmetry, no matter how far the system is from thermodynamic equilibrium. Specifically, given any cycle of reversible transitions, the ratio of the frequencies with which the cycle is traversed in one direction versus the other is determined, in the long-time limit, only by the thermodynamic force on the cycle itself, without requiring knowledge of transition rates elsewhere in the system. In particular, if there is no net energy expenditure on the cycle, then, over long times, the cycle traversal frequencies are the same in either direction. 
\end{abstract}

\section{INTRODUCTION}

Non-equilibrium systems are ubiquitous in nature, especially in biology. They are also notoriously difficult to deal with, lying largely beyond the scope of classical thermodynamics and the statistical mechanics developed in the 19th and early 20th centuries. In the 1960s, the biophysicist Terrell Hill introduced a diagrammatic method---essentially a graph---for analyzing individual biochemical entities, such as a membrane transporter complex, which operate stochastically under Markovian assumptions, away from thermodynamic equilibrium \cite{Hill_1966}. This approach was further developed in the 1970s in J\"urgen Schnakenberg's network theory \cite{Schnakenberg_1976}. One of the central ideas in their work was how cycles in the graph permitted macroscopic thermodynamic quantities, such as entropy production, to be related to stochastic mesoscopic quantities, such as fluxes between mesostates. Such systems could thereby be analyzed thermodynamically in spite of the global parametric complexity which arises away from thermodynamic equilibrium (below). For reasons that remain unclear, the use of such graphs then faded from sight for many years. They were not utilised for the major breakthroughs in non-equilibrium statistical mechanics which emerged in the 1990s in the work of Jarzynski, Crooks and others \cite{Jarzynski_1997,Crooks_1999,Gallavotti_Cohen_1995,Lebowitz_Spohn_1999}. More recently, as physicists have begun to build on these breakthroughs, the graph-based methods of Hill and Schnakenberg have came back into view \cite{Andrieux_Gaspard_2007,Rahav_Jarzynski_2007}, and the underlying Markov process representation has been widely adopted within the field of stochastic thermodynamics \cite{Seifert_2008,VdBroek_Esposito_2015}.

Independently of this physics tradition, an approach to timescale separation called the ``linear framework'' was introduced in the biological literature in 2012, based on directed graphs with labeled edges \cite{Gunawardena_2012}. The framework was originally applied to biochemical systems with large numbers of entities. For single entities, the approach may be seen, in a similar way to that of Hill and Schnakenberg, as a graph-based treatment of continuous-time, finite-state Markov processes. Vertices of the graph correspond to system states, edges to transitions between states and labels to infinitesimal transition rates. Such a graph is not merely a description of the system but a mathematical object in its own right, in terms of which system properties can be calculated, irrespective of how far the system is from thermodynamic equilibrium \cite{Wong_Gunawardena_2018b}. We will use the linear framework here to show how graph cycles can retain local sequence-reversal symmetry, despite the global complexity that emerges away from thermodynamic equilibrium. We hope these results will reinforce the significance of Hill and Schnakenberg's pioneering insights and draw further attention to the fertile area of study which lies between mathematics, physics and biology.

\section{THE LINEAR FRAMEWORK}
As the linear framework has been described in several publications, we outline here only what is needed to set our results in context; for more background and history, see \cite{Gunawardena_2012,Mirzaev_Gunawardena_2013}; for further details and relevant applications, see \cite{Estrada_2016,Wong_Gunawardena_2018,Wong_Gunawardena_2018b,Biddle_Nguyen_2019}; for reviews, see \cite{Gunawardena_2014b,Wong_Gunawardena_2019}. Most of the assertions below are well known in stochastic thermodynamics, if not always with the same notation and terminology, and justifications can be found in the references cited above. 

\begin{figure}[h]
\begin{center}
\includegraphics[width=0.95\textwidth]{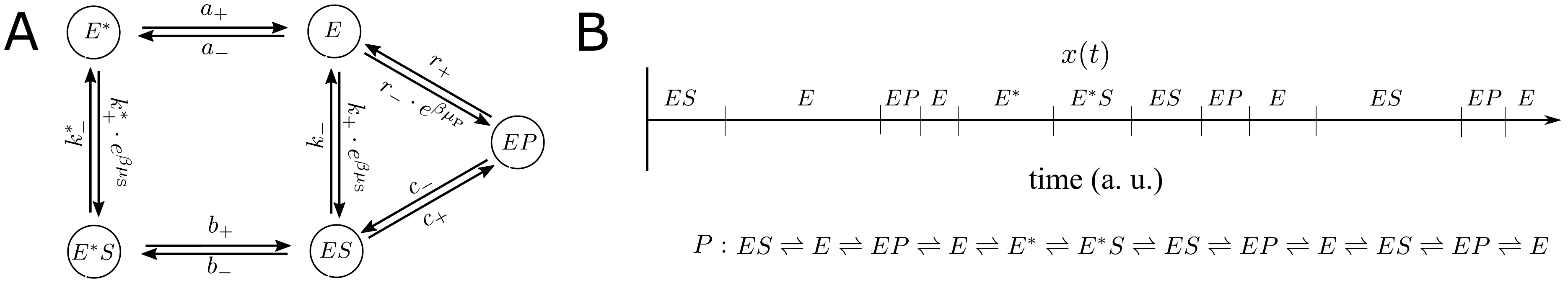}
\caption{\textbf{A}: A linear framework graph describing a variant of Ainslie et al.'s ligand-induced slow transition model \cite{Ainslie_1972} of the biological phenomenon of kinetic cooperativity or ``allokairy'' \cite{Hilser_2015}. An enzyme with two conformations, $E$ and $E^*$, converts substrate S to product P only in its active conformation, $E$. The system is in thermal contact with a thermally uniform environment at reciprocal temperature $\beta$, and is coupled to particle reservoirs for $S$ and $P$, each held at a characteristic chemical potential, $\mu_\mathrm{S}$ and $\mu_\mathrm{P}$, respectively. \textbf{B}: A realization of the underlying Markov process gives rise to a trajectory $x(t)$. The vertical marks signify the jumps between the mesostates annotated above. The corresponding path of reversible transitions is shown below. Trajectories convey temporal information and provide the dwell-time in each mesostate; paths convey only the order of transitions. Our results apply to paths rather than trajectories. 
\label{Sample-Graph}}
\end{center}
\end{figure}

A linear framework graph, $G$, is a finite, directed graph with labeled edges and no self-loops. For the purposes discussed here, the vertices (often denoted $1, \cdots, n$; the example in Fig. \ref{Sample-Graph} uses a more explanatory naming convention) represent the states in which one might find a given realization of a system. We call the vertices ``mesostates" to emphasize that while they refer to the configuration of the system and not to an ensemble, they are not microstates in the customary physics sense.  The edges, denoted $i \ra j$, represent transitions between mesostates; and the edge labels, denoted $\ell(i \ra j)$, represent infinitesimal transition rates for an underlying Markov process and are positive quantities with dimensions of [time]$^{-1}$. If we let $X(t)$ denote the underlying Markov process at time $t$, then this process is specified by a conditional probability distribution $\pr(X(t) = i \,|\, X(s) = j)$, where $i$ and $j$ are mesostates in $G$ and $s < t$. With this notation,
\begin{equation}
 \ell(i \ra j) = \lim_{\Delta t \ra 0} \ddfrac{\pr(X(t + \Delta t) = j \,|\, X(t) = i)}{\Delta t} \,. \label{DefRate}
\end{equation}
Edge labels may include expressions which specify the interaction between mesostates and entities in the environment of the graph, such as particle reservoirs or thermal energy (Fig. \ref{Sample-Graph}). For our purposes here, as we will explain below, we can consider the labels to be symbolic constants representing positive real numbers.

The master equation of the Markov process describes the time evolution of mesostate probabilities, described by the column vector, $\mathbf{p}(t)$, where $p_i(t)$ is the probability that the system is in mesostate $i$ at time $t$. This master equation can be obtained from $G$ as,
\[
\frac{dp_i}{dt}=\sum_{j\neq i} \big( p_j \ell(j \ra i) - p_i \ell(i \ra j) \big) \,.
\]
Note that, since each edge has only one source vertex, each $dp_i/dt$ is linear in the $p$'s.  We can therefore represent the master equation as a linear differential equation,
\begin{equation}
\frac{d\mathbf{p}}{dt} = \lap(G)\cdot\mathbf{p}(t), \label{Laplace}
\end{equation}
where $\lap(G)$ is the $n \times n$ Laplacian matrix of $G$ \cite{Gunawardena_2012,Mirzaev_Gunawardena_2013} and $\mathbf{p}$ is the column vector of mesostate probabilities. It is easy to see that the column sums of $\lap(G)$ are zero, which corresponds to the conservation of total probability. The relationship described above between graphs and Markov processes is quite general: given any continuous-time, finite-state Markov process, for which infinitesimal rates can be defined, there is a graph $G$ for which Eq. \ref{Laplace} specifies the master equation \cite{Mirzaev_Gunawardena_2013}.  

For the applications considered here, we assume that the environmental variables affecting the rates do not change on the timescale over which the system is studied, and that temperatures and chemical potentials of the reservoirs are unaffected by any transitions that the system might undergo.  We therefore treat the edge labels as symbolic constants.

This graph-based description of the Markov process allows effective calculation of quantities of interest whether or not the system is at thermodynamic equilibrium (Eqs. \ref{NEQ-SS} and \ref{e-prs} below). We outline the details of how this is done in the next section (section \ref{prelim}) before stating and proving our main results (section \ref{main}). 

Fig. \ref{Sample-Graph} gives an example of a linear framework graph used to model a biochemical system.  We refer to this example to explain our approach. In section \ref{sims} we present Gillespie simulations of the underlying system as an illustration of our results and how they might be employed experimentally.

\section{PRELIMINARY RESULTS \label{prelim}}

\subsection*{Departure from equilibrium in the linear framework}

\subsubsection*{Steady states and the Matrix-Tree theorem}

\begin{figure}[b]
\begin{center}
\includegraphics[width=0.65\textwidth]{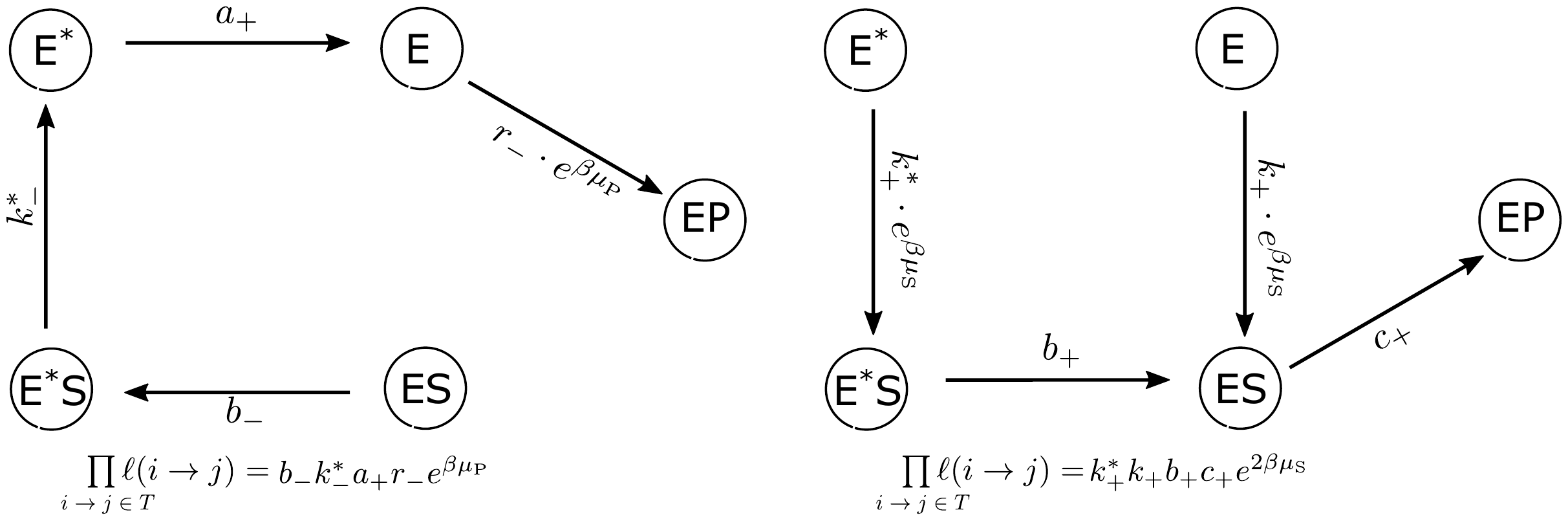}
\caption{Two of the 11 spanning trees rooted at vertex $EP$.  Each tree contributes a term to the expression in Eq. \ref{NEQ-SS}, as shown below each tree. \label{Sample-Trees}}
\end{center}
\end{figure}

We will assume from now on that $G$ is strongly connected, so that any two vertices, $i, j$, are connected by a directed path of continguous edges, $i = i_1 \ra \cdots \ra i_k = j$. In this case, Eq. \ref{Laplace} has a unique steady state, $\mathbf{p}^*$. Equivalently, the kernel of the Laplacian matrix is one-dimensional, $\dim\ker\lap(G) = 1$. A canonical basis element, $\rho(G) \in \ker\lap(G)$, can be calculated from $G$ through the Matrix-Tree theorem (MTT), 
\begin{equation}
\rho_i(G) = \sum_{T \in \Theta_i(G)} \left( \prod_{j \ra k \in T} \ell(j \ra k) \right) \,.
\label{NEQ-SS}
\end{equation}
Here, $\Theta_i(G)$ is the set of spanning trees of $G$ rooted at vertex $i$. A spanning tree is a subgraph of $G$ which includes all vertices of $G$ (spanning) and has no cycles when edge directions are ignored (tree). It is rooted at $i$ if $i$ is the only vertex with no outgoing edges. Fig. \ref{Sample-Trees} shows two spanning trees rooted at the mesotate $EP$ for the graph in Fig. \ref{Sample-Graph}A. 

Since $\mathbf{p}^* = \lambda\rho(G)$, for some scalar $\lambda \in \R^{+}$, it follows from the conservation of probability that,
\begin{equation}
p^*_i = \frac{\rho_i(G)}{\rho_1(G) + \cdots + \rho_n(G)} \,.
\label{e-prs}
\end{equation}
Eq. \ref{e-prs} expresses steady-state probabilities in terms of the edge labels, through Eq. \ref{NEQ-SS}, and can be applied to a system maintained arbitrarily far from thermodynamic equilibrium. The MTT is not required for our main result but, in conjunction with Eq. \ref{e-zss} below, it illustrates the dramatic increase in global parametric complexity that arises away from equilibrium. Eq. \ref{NEQ-SS} shows that the steady-state probability of each mesostate becomes dependent in a complicated way on edge labels throughout the graph. The MTT, first proved in this form by Tutte \cite{Tutte_1948}, was known to Hill \cite{Hill_1966} and Schnakenberg \cite{Schnakenberg_1976} but then seems to have disappeared from view; for its convoluted history across many disciplines and a proof, see \cite{Mirzaev_Gunawardena_2013}.

\subsubsection*{Thermodynamic equilibrium}

If a system can reach thermodynamic equilibrium, then it must obey the principle of detailed balance. This principle holds at all scales from elementary to mesoscopic systems: observing a trajectory over time, transitions from configuration $i$ to $j$ and from configuration $j$ to $i$ will happen with equal frequency, however coarse- or fine-grained the conceptions of $i$ and $j$ may be \cite{Ter_Haar_1954}. Detailed balance arises from time-reversal symmetry of the underlying fundamental laws of physics \cite{Mahan_1975}.  

Let us denote the flux along the edge $i \ra j$, or $p_i\ell(i \ra j)$, by $J_{i\to j}$; this is proportional to the average rate at which the system makes the transition $i \to j$. In order to satisfy detailed balance, the graph must be reversible: given the edge $i \ra j$, there must exist an edge $j \ra i$, which represents the reverse transition (and not merely an alternative way of returning from $j$ to $i$). It must also be the case that the forward and reverse fluxes equal each other, so that the net flux vanishes, $J_{i \ra j} - J_{j \ra i} = 0$. Equivalently, 
\begin{equation}
\frac{p^*_j}{p^*_i}=\frac{\ell(i \to j)}{\ell(j\to i)}. \label{DB-graph}
\end{equation}
For systems that can reach equilibrium in contact with one or more reservoirs (of particles, heat, etc), a free energy, $F(i)$ can be associated with each mesostate, $i$. and this gives a physical interpretation for the ratios of transition rates,
\begin{equation}
\log\left(\frac{\ell(i \to j)}{\ell(j \to i)}\right) = e^{\beta \Delta F_{ij}} \,,
\label{StatMechTransitions}
\end{equation}
where $\Delta F _{ij} = F(i) - F(j)$. This identification goes back to Hill \cite{Hill_1966} and Schnakenberg \cite{Schnakenberg_1976} and we discuss below (Eq. \ref{LDB}) how it has been extended to non-equilibrium systems. 

Eq. \ref{DB-graph} leads to an alternative basis element $\mu(G) \in \ker\lap(G)$. Choose a reference mesostate in $G$, usually taken to be the mesostate indexed by $1$. Given any path of reversible edges from $1$ to a given mesostate $i$, $1 = i_1 \rla i_2 \rla \cdots \rla i_{k-1} \rla i_k = i$, let
\begin{equation}
\mu_i(G) = \left(\frac{\ell(i_1 \ra i_2)}{\ell(i_2 \ra i_i)}\right) \cdots \left(\frac{\ell(i_{k-1} \ra i_k)}{\ell(i_k \ra i_{k-1})}\right) \,.
\label{e-zss}
\end{equation}
It is clear from Eq. \ref{DB-graph} that $\mu_i(G)$ does not depend on the chosen path from $1$ to $i$ and that, because $\dim\ker\lap(G) = 1$ and $\mu_1(G) = 1$, $p^*_1\mu(G) = p^*_i$. It follows that $\mu(G) \in \ker\lap(G)$ and that $\rho_1(G)\mu(G) = \rho(G)$.  The denominator of Eq. \ref{e-prs} is then proportional to the partition function for the relevant ensemble of statistical mechanics. It also provides an analog of the partition function for a non-equilibrium system.   

\subsubsection*{Non-equilibrium complexity and the breakdown of detailed balance}

On the other hand, a system may be maintained away from thermodynamic equilibrium.  The comparison between $\mu(G)$, at thermodynamic equilibrium, and $\rho(G)$, away from equilibrium, is instructive. With the former, Eq. \ref{e-zss} shows that it is sufficient to take any single path in $G$ to a given mesostate to calculate its steady-state probability using Eq. \ref{e-prs}. With the latter, when detailed balance is broken, not only do steady-state probabilities become path dependent, every path in $G$ makes a contribution to them and the MTT in Eq. \ref{NEQ-SS} provides the bookkeeping for this calculation. The combinatorial explosion coming from enumerating all rooted spanning trees can be super-exponential in the size of $G$ \cite{Estrada_2016} and the steady-state probability of a mesostate can come to depend in an extremely complicated manner on all the labels in $G$. The graph-based approach here provides a vivid demonstration of the profound difference between equilibrium and non-equilibrium states. 

The impact of non-equlibrium path-dependency is felt globally even if energy expenditure is limited and local. If we take a graph $G$ that obeys detailed balance, and perturb even a single edge label, $\ell(i \ra j)$, so as to break detailed balance, the resulting change in $\mathbf{p}^*$ due to Eq. \ref{NEQ-SS} leads to a widespread breakdown of detailed balance (Eq. \ref{DB-graph}) even at edges whose labels retain their equilibrium values. This point is clearly illustrated in the simulations undertaken below (section \ref{sims}) of the example in Fig. \ref{Sample-Graph}A.  

A system might be maintained away from equilibrium through contact with multiple reservoirs at different temperatures or chemical potentials. This is common in biological systems, from molecular motors which exploit a chemical potential difference between ATP, on the one hand, and ADP and inorganic phosphate, $P\textsubscript{i}$, on the other, to enzymes which operate in the presence of a chemical potential difference between substrate and product, as in the example in Fig. \ref{Sample-Graph}. Alternatively, chemical or other energy may be consumed by the system itself and then dissipated to the environment, as in the case of active matter. We note that either kind of non-equilibrium system can be treated within the linear framework.

\subsection*{Cycles and affinities}

An equivalent condition to the statement of detailed balance in Eq. \ref{DB-graph} can be given in terms only of the edge labels. Consider any cycle of reversible edges, $i_1 \rla i_2 \rla \cdots \rla i_{m-1} \rla i_m = i_1$. Cycles need not be simple: the same mesostate may appear more than once, with more than one index, in the numbered list. At thermodynamic equilibrium, the product of the labels in one direction around the cycle equals the product in the other direction, 
\begin{equation}
\ell(i_1 \ra i_2) \cdots \ell(i_{m-1} \ra i_m) = \ell(i_m \ra i_{m-1}) \cdots \ell(i_2 \ra i_1) \,.
\label{e-cyc}
\end{equation}
Eq. \ref{e-cyc} makes clear that, if $G$ can reach thermodynamic equilibrium, the edge labels are not independent quantities. The label ratios on the edges of any rooted spanning tree form a set of independent parameters, in terms of which all other label ratios can be determined using Eq. \ref{e-cyc} \cite{Estrada_2016}. If a graph is at thermodynamic equilibrium, flux balance as given by Eq. \ref{DB-graph} and the cycle condition as given by Eq. \ref{e-cyc} are equivalent statements \cite{Biddle_Nguyen_2019}.

For any cycle $C$ as described above, there is an affinity $\tilde{A}(C)$ associated with the cycle, defined as the logarithm of the ratio of the corresponding terms in Eq. \ref{e-cyc},
\begin{equation}
\tilde{A}(C) = \log\left(\frac{\ell(i_1 \ra i_2) \cdots \ell(i_{m-1} \ra i_m)}{\ell(i_2 \ra i_1) \cdots \ell(i_m \ra i_{m-1})}\right) \,.
\label{AffDef1}
\end{equation}
It follows from Eq. \ref{e-cyc} that for a system at thermodynamic equilibrium $\tilde{A}(C) = 0$ for all cycles. We follow Schnakenberg \cite{Schnakenberg_1976} and Hill \cite{Hill_1966} in this definition of the affinity.

The affinity of a cycle has a natural thermodynamic interpretation, which arises from an extension of Eq. \ref{StatMechTransitions} to systems maintained away from equilibrium. For a reversible edge, $i \rla j$, 
\begin{equation}
\log\left(\frac{\ell (i \to j)}{\ell(j \to i)}\right) = \Delta S^\mathrm{tot}_{ij} \,.
\label{LDB}
\end{equation}
Here, the right-hand side represents the total change in entropy of the system and all reservoirs when the system makes the transition from $i$ to $j$. Eq. \ref{LDB} is referred to as ``local detailed balance'', which can be justified under a broad range of conditions \cite{Bauer_Cornu_2015}.  To take an example from the graph in Fig. \ref{Sample-Graph}A, the pair of substrate binding and unbinding transitions would satisfy,
\begin{align*}
\log\left(\frac{\ell(E \to ES)}{\ell(ES \to E)}\right) &= \beta \left(F(E) - F(ES) + \mu_\mathrm{S}\right).
\end{align*} 
It follows readily from Eq. \ref{LDB} that the affinity of a cycle is given by,  
\begin{equation}
\tilde{A}(C) = \Delta S^\mathrm{res} \,,
\label{LDB-cycle}
\end{equation}
where $\Delta S^\mathrm{res}$ is the total entropy produced in the reservoirs by one completion of the cycle \cite{Hill_1975}.

In considering the affinity, $\tilde{A}(C)$, of cycle $C$, it is helpful to keep in mind that only minimal cycles contribute to it. A cycle is minimal if it contains no repeated mesotates and at least three mesostates. For instance, for the graph in Fig. \ref{Sample-Graph}A, the cycle $E \rla ES \rla E$ is not minimal. It is a simple instance of an ``excursion'' which takes a path from $E$ to $ES$ and then returns along the reverse path. There is no net energy expenditure along such excursions---in this case a substrate molecule, $S$, is taken from its reservoir and then returned to it---and the affinity is zero. The only parts of a cycle which contribute to the affinity are those corresponding to minimal cycles. 

As we have defined it, a cycle specifies its starting mesostate, but the cycle properties described above are independent of the starting point. The following notation will therefore be helpful. Given a cycle $C:\,\, i_1 \rla i_2 \rla \cdots \rla i_{m-1} \rla i_m = i_1$, let $C(i_j)$, where $1 \leq j \leq m$, denote the cycle with the same order of transitions starting at mesosate $i_j$, $C(i_j):\,\, i_j \rla i_{j+1} \rla \cdots \rla i_{m-1} \rla i_1 \rla i_2 \rla \cdots \rla i_{j-1} \rla i_j$. It follows that $C(i_1) = C(i_m) = C$. Let $\{C\}$ denote the set of cycles having the same cyclic order of transitions, $\{C\} = \{C(i_1), C(i_2), \cdots, C(i_{m-1})\}$. Note that the cycle condition in Eq. \ref{e-cyc} and the affinity in Eq. \ref{AffDef1} are both well-defined as properties of $\{C\}$.

\section{MAIN RESULT: REVERSAL SYMMETRY OF CYCLES \label{main}}

We need some further notation to state our main result. Let $P:\,\, i_1 \rla i_2 \rla \cdots \rla i_{m-1} \rla i_m$ be any path of reversible edges in a graph $G$, where $m > 2$.  Consider observing a trajectory of the underlying Markov process (Fig. \ref{Sample-Graph}B). Denote the number of times that $P$ occurs on the trajectory by $n[P,x(t)]$. It is important that all the mesostates on the path are observed in the specified order without deviations. For the example trajectory in Fig. \ref{Sample-Graph}B and for the path $P: ES \rla EP \rla E$, $n[P,x(t)]=2$, while for the cycle $P: E \rla ES \rla EP \rla E$, $n[P,x(t)] = 1$. 

Let $\pr(P)$ denote the probability that the Markov process, having reached the initial mesostate $i_1$, will subsequently go through the transitions in $P$ in exactly the specified order without deviation. Since $G$ is strongly connected with positive labels, we can appeal to the ergodic theorem for Markov processes \cite{Stroock_2005} to interpret the probability in frequentist terms, 
\begin{equation}
\pr(P) = \lim_{t \ra \infty} \frac{n[P,x(t)]}{n[i_1,x(t)]} \,,
\label{e-freqi}
\end{equation}
where the denominator is the number of times in which the starting mesostate, $i_1$, is observed along the trajectory.

Given a cycle $C:\,\, i_1 \rla i_2 \rla \cdots \rla i_{m-1} \rla i_m = i_1$ let $C^r$ denote the reverse cycle, $C^r:\,\, i_m \rla i_{m-1} \rla \cdots \rla i_2 \rla i_1 = i_m$. We can easily extend the definition to cycle sets by defining $\{C\}^r = \{C^r\}$. Along a trajectory, the same cyclic order of transitions may be observed starting at any mesostate, so we can define $n[\{C\},x(t)]$ to be,
\[ n[\{C\},x(t)] = n[C(i_1),x(t)] + \cdots + n[C(i_{m-1}),x(t)] \,.\]

\vspace{3mm}
\noindent
\textbf{Theorem:} For any connected reversible graph, any cycle $C:\,\, i_1 \rla i_2 \rla \cdots \rla i_{m-1} \rla i_m = i_1$ of reversible edges in $G$ and any mesostate $i_j$ on $C$, 
\[ \lim_{t \ra \infty} \frac{n[C(i_j),x(t)]}{n[C(i_j)^r,x(t)]} = \lim_{t \ra \infty} \frac{n[\{C\},x(t)]}{n[\{C^r\},x(t)]} = e^{\tilde{A}(\{C\})}. \]
\vspace{1 mm}

The proof of this requires a couple of steps. The first step is to calculate probabilities. Under the conditions of the theorem, we claim that
\begin{equation}
\ddfrac{\pr(C(i_j))}{\pr(C(i_j)^r)} = e^{\tilde{A}(C)} \,.
\label{e-aec}
\end{equation}
To see this, note that, conditional on the system being in mesostate $i_1$, the probability that its next transition will be to state $i_2$ is given by
\begin{equation}
\ddfrac{\ell(i_1 \to i_2)}{\sum_{i_1 \to j}\ell(i_1 \to j)}, \label{Prob-l-1}
\end{equation}
where the sum in the denominator is over all outgoing edges in $G$ from $i_1$. Let us denote this latter quantity, for a given mesostate $i$, by $\Z(i) = \sum_{i \ra j}\ell(i \ra j)$, so that $\Z(i_1)$ gives the denominator in Eq. \ref{Prob-l-1}. Since the probability in Eq. \ref{Prob-l-1} leads to mesostate $i_2$, the calculation can be continued to determine the probability, conditional on starting in mesostate $i_1$, that the system makes the transition from $i_1$ to $i_2$ followed by the transition from $i_2$ to $i_3$, 
\begin{equation*}
\frac{\ell(i_1 \to i_2)}{\mathcal{Z}(i_1)}\cdot\frac{\ell(i_2 \to i_3)}{\mathcal{Z}(i_2)}.
\end{equation*}
It follows by induction that, conditional on starting in mesostate $i_1$, the probability that the next $m-1$ transitions will be as specified by cycle $C$ is given by, 
\begin{equation}
\ddfrac{\prod_{k=1}^{m-1} \ell(i_k \to i_{k+1})}{\prod_{k=1}^{m-1} \Z(i_k)}. \label{PrCa}
\end{equation}
The equivalent expression for $C^r$ is given by, 
\begin{equation}
\ddfrac{\prod_{k=2}^m \ell(i_k \to i_{k-1})}{\prod_{k=2}^m \Z(i_k)}. \label{PrCa-prime}
\end{equation}
Hence, taking the ratio of Eq. \ref{PrCa} by Eq. \ref{PrCa-prime} and recalling that $i_1 = i_m$ and the definition of affinity in Eq. \ref{AffDef1}, we see that, 
\begin{equation}
\ddfrac{\pr(C)}{\pr(C^r)} = \left(\ddfrac{\Z(i_m)}{\Z(i_1)}\right)\ddfrac{\prod_{k=1}^{m-1} \ell(i_k \to k_{k+1})}{\prod_{k=2}^m \ell(i_k \to i_{k-1})} = e^{\tilde{A}(C)}.
\end{equation}
This final value is independent of the starting mesostate $i_1$, from which the claim made in Eq. \ref{e-aec} follows.

We can now use Eq. \ref{e-freqi} to interpret probabilities in terms of cycle counts. We see that 
\[ \pr(C(i_j)) = \lim_{t \ra \infty} \frac{n[C(i_j),x(t)]}{n[i_j,x(t)]} \hspace{1em}\mbox{and}\hspace{1em} \pr(C(i_j)^r) = \lim_{t \ra \infty} \frac{n[C(i_j)^r,x(t)]}{n[i_j,x(t)]} \,.\] 
It follows from Eq. \ref{e-aec} that both of these quantities are non-zero and, therefore, 
\begin{equation}
\lim_{t \ra \infty} \frac{n[C(i_j),x(t)]}{n[C(i_j)^r,x(t)]} = \lim_{t \ra \infty} \frac{n[C(i_j),x(t)]/n[i_j,x(t)]}{n[C(i_j)^r,x(t)]/n[i_j,x(t)]} = \frac{\pr(C(i_j))}{\pr(C(i_j)^r)} = e^{\tilde{A}(C)} \,.
\label{e-ncij}
\end{equation}
This establishes part of the theorem. 

The remaining part comes from the following observation, whose proof is elementary. If functions $a_j(t)$ and $b_j(t)$ are defined for $1 \leq j \leq m$ and satisfy $\lim_{t \ra \infty}a_j(t)/b_j(t) = \alpha$ for each $j$, then
\begin{equation}
\lim_{t \ra \infty}\frac{a_1(t) + \cdots + a_m(t)}{b_1(t) + \cdots + b_m(t)} = \alpha \,.
\label{e-limt}
\end{equation}
Taking $a_j(t) = n[C(i_j),x(t)]$ and $b_j(t) = n[C(i_j)^r,x(t)]$ and using Eq. \ref{e-ncij} completes the proof of the theorem. 

\section{SIMULATIONS \label{sims}}

\begin{figure}[!h]
\center
\includegraphics[width = 0.8\textwidth]{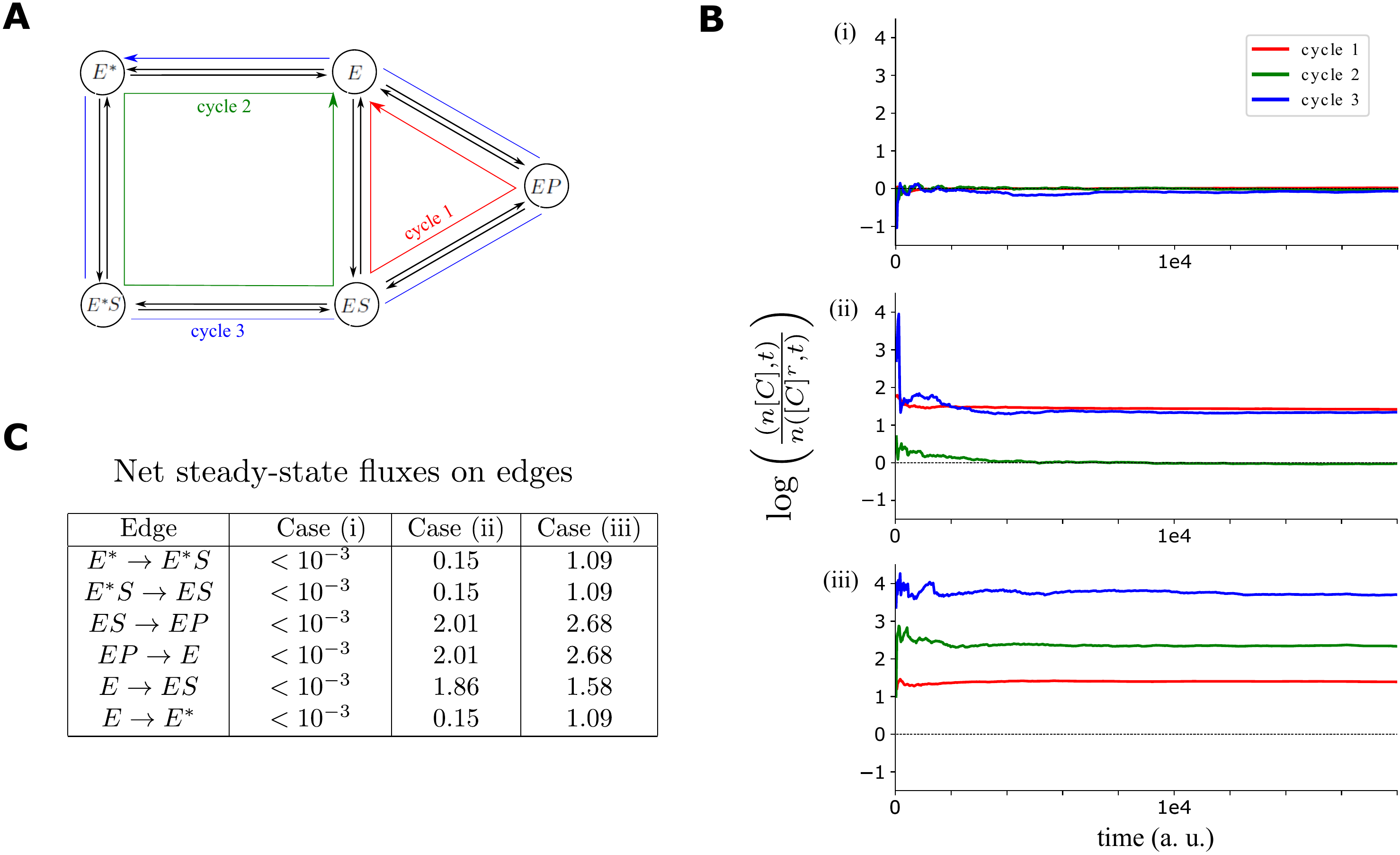}
\caption{Illustration of symmetry properties of cyclic paths.  \textbf{A} The example graph from Fig. \ref{Sample-Graph} is shown with three cycles color coded. The text explains the three scenarios which were studied. \textbf{B.} Logarithmic plots of the reversibility ratio in the theorem, with time, $t$, in arbitrary units, for each scenario, as annotated to the top left of each plot. \textbf{C.} The net flux, $J_{i \ra j} - J_{j \ra i}$ associated with each pair of forward and reverse transitions, $i \rla j$, is shown for each scenario. The first column gives the edge $i \ra j$ in the direction of net flux.  (The values for case (i) denote absolute values). The text discusses the flux patterns. \label{f-1}}
\end{figure}

Fig. \ref{f-1} illustrates the properties of cyclic paths for the allokairy example in Fig. \ref{Sample-Graph}. Three minimal cycles (section \ref{prelim}) are shown color coded in Fig. \ref{f-1}A. The system was simulated for three scenarios. Case (i): thermodynamic equilibrium, with $\mu_\mathrm{S} = \mu_\mathrm{P}$. Case (ii): the system is driven from equilibrium by a difference in chemical potential between substrate and product, $\mu_\mathrm{S} > \mu_\mathrm{P}$, so that the reaction favors product formation. The only difference from the equilibrium case is a difference in the rate of product re-binding, for the edge $E \ra EP$. Case (iii): two different forces drive the system away from equilibrium.  As in case (ii), $\mu_\mathrm{S} > \mu_\mathrm{P}$.  Additionally, rather than merely being different conformations, $E^*$ and $E$ are treated as phosphorylated and non-phosphorylated forms of the enzyme, respectively. The process of phosphorylation and dephosphorylation is driven by ATP hydrolysis, with the system in contact with reservoirs of ATP, ADP and P\textsubscript{i}. Cycles 2 and 3 experience the difference in chemical potential given by $\mu_\mathrm{ATP}-\mu_\mathrm{ADP}-\mu_\mathrm{P_i}$. The temperature and chemical potentials used in the simulations, as well as the transition rates themselves, are provided as Supplemental Information. 

We carried out Gillespie simulations for this model in the three different scenarios and recorded the resulting edge fluxes and cycle occurrence statistics.  In case (i), the system is at thermodynamic equlibrium. It is coupled to two particle reservoirs, of $S$ and $P$, whose chemical potentials are equal: $\mu_\mathrm{S} = \mu_\mathrm{P}$.  In this case, all net fluxes vanish (Fig. \ref{f-1}C, column 2) and all cycle reversiblity ratios tend to unity (Fig. \ref{f-1}B, top). In case (ii), $\mu_\mathrm{P} < \mu_\mathrm{S}$ so that the reaction favors product formation. Cycles 1 and 3 thus have affinity $\tilde{A}(C)=\beta(\mu_\mathrm{S} - \mu_\mathrm{P})$, while cycle 2 has zero affinity (Fig. \ref{f-1}B, middle). In case (iii), cycles 2 and 3 are driven by ATP-dependent phosphorylation. The affinity of cycle 3 becomes $\tilde{A}(C_3) = \beta\left(\mu_\mathrm{ATP} + \mu_\mathrm{S} - \mu_\mathrm{ADP} -\mu_\mathrm{P_i} - \mu_\mathrm{P} \right)$, and cycle 2 acquires a nonzero affinity as well, $\tilde{A}(C_2) = \beta\left(\mu_\mathrm{ATP}-\mu_\mathrm{ADP}-\mu_\mathrm{P_i}\right)$. In both non-equilibrium scenarios all reversible edges in the graph acquire net fluxes (Fig. \ref{f-1}C, columns 3 \& 4). At steady state, the net flux from $E$ to $E^*$, from $E^*$ to $E^*S$ and from $E^*S$ to $ES$ must be equal, as must the net flux from $ES$ to $EP$ and from $EP$ to $E$, while the net flux from $E$ to $ES$ gives the difference between these two amounts (in case (iii) the slight discrepancy in flux balance arises from rounding error).

Case (ii) differs from the equilibrium case (i) only in one rate, $\ell(E \ra EP)$. Nevertheless, net fluxes appear at all reversible edges. In fact, cases (ii) and (iii) are difficult to distinguish by an examination of net fluxes alone. The reversal ratio plots, however, look quite different. In case (i), symmetry is preserved for all cycles.  In case (ii), cycle 2 retains its reversal symmetry, while it is broken for cycles 1 and 3 to the same extent. In case (iii), reversal symmetry is broken for all cycles, each to a different extent.

\section{DISCUSSION \label{disc}}

The principle of detailed balance at equilibrium is stated succinctly by Dirk ter Haar: `` ...at equilibrium the number of processes destroying situation A and creating situation B will be equal to the number of processes producing A and destroying situation B'' \cite{Ter_Haar_1954}. Hence, the observation of any process happening more frequently than its reverse is confirmation that the system is away from equilibrium.  However, as Fig. \ref{f-1}C clearly shows, the breakdown of this symmetry in a non-equilbirum system is so pervasive that a single transition that is driven away from equilibrium can disrupt the balance between every pair of forward and reverse transitions. Furthermore, it can make the steady-state probability of every mesostate dependent on all the rates in the system. Even for a simplified system like the one in Fig. \ref{Sample-Graph}A, Eq. \ref{NEQ-SS} shows that one must enumerate 55 spanning trees (Fig. \ref{Sample-Trees}), each containing four factors, to compute the non-equilibrium steady-state probabilities of the system.  By contrast, the probabilities can be computed at equilibrium from four ratios of rates.  Notwithstanding this path-dependent complexity, cyclic paths on which there is no net energy expenditure retain a remarkable symmetry: for long trajectories, one can expect to observe such a cycle and its reverse with the same frequency. Furthermore, the extent to which this symmetry is broken for a cycle which is driven away from equilibrium is a function only of the affinity of the cycle, irrespective of thermodynamic forces acting elsewhere in the system. 

As noted in the Introduction, the use of graph cycles to study the thermodynamics of Markov processes, goes back to the work of Hill and Schnakenberg. Schnakenberg, following Hill, introduced a measure of cycle flux, which, together with the cycle affinity (Eq. \ref{AffDef1}), allows the overall entropy production to be determined \cite{Schnakenberg_1976}. These cycle fluxes are distinct from the cycle counts considered here. Hill introduced a way of counting cycles along a trajectory and analyzed the fluctuations in this quantity \cite{Hill_1975,Hill_Chen_1975}. Jiang, Qian and Qian subsequently provided rigorous proofs of Hill's observations as well as several other results \cite{JiangDa-Quan2004Mton}. The cycle count definitions in these papers are different from ours. Hill and Jiang \emph{et al} focus only on minimal cycles, as defined in section \ref{prelim}, and they count cycle completions, allowing for excursions away from the cycle, by recursively removing cycles from a path. For them, the trajectory in Fig. \ref{Sample-Graph}B contains only the completed cycles $ES \rla E \rla E^* \rla E^*S \rla ES$ and $ES \rla EP \rla E \rla ES$. The resulting asymptotic counts over long trajectories are entirely appropriate for determining fluxes on edges and calculating overall entropy production. If the asymptotic count for a cycle is equal to the asymptotic count for the reversed cycle, and this is true for all minimal cycles in the graph, then the system is at thermodynamic equilibrium \cite[Theorem 2.2.10]{JiangDa-Quan2004Mton}. 

In contrast, we consider any cycle and our way of counting cycle occurrences, rather than cycle completions, does not allow excursions away from the cycle. Accordingly, we find the second cycle given above in the trajectory in Fig. \ref{Sample-Graph}B, counted once with the given starting mesostate.  We do not count any occurrences of the first cycle given above but we do count single occurrences of four other cycles: $EP\rla E\rla E^* \rla E^*S\rla ES \rla EP$, $E\rla E^* \rla E^*S\rla ES \rla EP \rla E$, $EP \rla E \rla ES \rla EP$, and $E \rla ES \rla EP \rla E$. The asymptotic ratio of forward cycle count compared to reversed cycle count provides local information about net energy expenditure on that cycle, irrespective of whether or not there is energy expenditure elsewhere in the system.

The behavior under sequence reversal of our cycle counts, as given by the theorem, is suggestive of a symmetry. Non-equilibrum fluctuation theorems in the Markov process setting reveal a striking symmetry in the large deviation function, or cumulant generating function, for asymptotic fluctuations in stochastic quantities like fluxes (``currents'') or entropy production \cite{Lebowitz_Spohn_1999,Andrieux_Gaspard_2007,Garcia_2012,Wachtel_2018}. The reversal symmetry we find pertains only to the mean of the cycle occurrence distribution. It is conceivable that this could be recovered in some way from the more general symmetries of fluctuation theorems but we know of no way to do so at present.

As Hopfield first showed for the case of error correction in the synthesis of biological macromolecules, thermodynamic equilibrium imposes fundamental limits---we have called them Hopfield barriers \cite{Estrada_2016}---on the functional capacities of biological systems \cite{Hopfield_1974}. These limits can be exceeded only by the expenditure of energy. The need to account explicitly for the non-equlilibrium nature of biological processes has been felt even more acutely in recent years, and understanding the implications of the breakdown of detailed balance has been central to this effort \cite{Tu_2008,Estrada_2016,Biddle_Nguyen_2019}. While detecting and measuring departure from equilibrium has become feasible in systems where mechanical work is done, as in observations of detailed-balance violations in the dynamics of bacterial flagella \cite{Battle_MacKintosh_2016}, it has proved more challenging for cellular information processing \cite{Liu_Wang_2020}. 

Cycle counting offers an alternative approach for analyzing non-equilibrium behavior which is becoming experimentally feasible. Advances in single-molecule experimental techniques are making it possible to observe in real time such processes as the conformational and binding behavior of enzymes \cite{Lu_2013}, the processivity of molecular motors \cite{Sikor_Lamb_2013}, and the binding and unbinding of transcription factors during gene regulation \cite{Chen_Liu_2014}. These developments hold out the tantalizing possibility that the results presented here could be used not only to confirm departure from thermodynamic equilibrium but also to identify the sources and quantify the extent of energy expenditure. 

\subsection*{Acknowledgements}
The authors thank two anonymous reviewers for their constructive suggestions which led to several improvements in the exposition and Jeremy Owen and Chris Jarzynski for helpful comments on the results. 
\bibliographystyle{ieeetr}

\end{document}